\documentclass[prl,reprint,twocolumn,preprintnumbers,amsmath,amssymb,showpacs,footinbib,superscriptaddress]{revtex4-1}

\usepackage[titletoc,toc,title]{appendix}
\usepackage{braket}
\usepackage[final]{feynmp}
\usepackage{ifpdf}
\usepackage{comment}
\usepackage{amsmath}
\usepackage{mathrsfs}
\usepackage{color}
\usepackage{bm}
\usepackage[font=small]{subcaption}
\usepackage[font=small]{caption}

\graphicspath{{./Figures/}}

\makeatletter
\def\endfmffile{%
	\fmfcmd{\p@rcent\space the end.^^J%
			end.^^J%
			endinput;}%
	\if@fmfio
		\immediate\closeout\@outfmf
	\fi
	\ifnum\pdfshellescape=\@ne
		\immediate\write18{mpost \thefmffile}%
	\fi}
\makeatother

\usepackage[demo]{graphicx}
\usepackage{dcolumn}
\usepackage{bm}
\usepackage{hyperref}

\newcommand{\beq}{\begin{equation}}
\newcommand{\eeq}{\end{equation}}

\def\8{\infty}
\def\oh{\frac{1}{2}}

\def\undertext#1{\vtop{\hbox{#1}\kern 1pt \hrule}}

\def\pp#1{\frac{\partial}{\partial#1}}
\def\pbyp#1#2{\frac{\partial#1}{\partial#2}}

\def\be{\begin{equation}}
\def\ee{\end{equation}}
\def\bea{\begin{eqnarray} & &}
\def\eea{\end{eqnarray}}

\def\rf#1{(\ref{#1})}

\begin{document}


\title{Singularities in the Loschmidt echo of quenched  topological superconductors}

\author{Sankalp Gaur}
\author{Victor Gurarie}
\affiliation{Department of Physics and Center for Theory of Quantum Matter, University of Colorado, Boulder, Colorado 80309, USA}
\author{Emil A. Yuzbashyan}
\affiliation{Department of Physics and Astronomy, Center for Materials Theory, Rutgers University, Piscataway, NJ 08854 USA}

\begin{abstract}
We study the Loschmidt echo in the quenched two-dimensional  $p$-wave topological superconductor. We find that if this superconductor is quenched out of the critical point separating its topological  and non-topological phases into
either of the two gapful phases, its Loschmidt echo features singularities occurring periodically in time where the second derivative of the Loschmidt echo over time diverges logarithmically. Conversely, we give arguments towards $s$-wave 
superconductors not having singularities in their Loschmidt echo regardless of the quench.   We also demonstrate that the conventional mean field theory calculates classical echo instead of its quantum counterpart, and  show how it should be
modified to capture the full quantum Loschmidt echo.  \end{abstract}

\maketitle


About a decade ago Loschmidt echo was proposed as yet another way of characterizing the dynamics of quenched quantum 
systems~\cite{Kehrein2013}.  Consider a quantum system residing in a state $\left| \Psi_i \right\rangle$, e.g., in the 
ground state.  Suppose the system Hamiltonian suddenly changes triggering  time evolution. After some time $t$ the time-evolved state is projected back onto $\left| \Psi_i \right\rangle$. The Loschmidt echo is defined as
\footnote{The standard definition of the Loschdmidt echo is 
$\left| \left< \Psi_i \right| e^{i H_2 t} e^{-i H_1 t}
\left| \Psi \right> \right|^2$. If $\Psi$ is the eigenstate of $H_1$, up to taking its absolute value square it effectively reduces to Eq.~\rf{eq:Loschmidt}.}
\be \label{eq:Loschmidt}  {\cal Z}(t) = \left\langle \Psi_i \right| e^{-i \hat H t} \left| \Psi_i \right\rangle. 
\ee
Of particular interest is the question of whether ${\cal Z}(t)$ may be non-analytic as a function of $t$, in a way reminiscent of the thermal partition function of quantum systems being non-analytic when they undergo phase transitions. 

This question was explored in the literature for a wide variety of quantum systems. Here we would like to demonstrate that a two-dimensional (2D) $p$-wave superconductor when quenched out of a critical point separating its two phases \cite{Read2000} (BEC, or strongly paired, or non-topological phase and BCS, or weakly paired, or topological phase) into either of its two phases features the Loschmidt echo with periodically occurring singularities. The singularities in $  {\cal Z}(t)$ are of the kind $(t-t_n)^2 \ln \left| t-t_n \right|$,  where  
  $t_n = (n+1/2) \pi/\left| \xi_0 \right|$ and $\left| \xi_0 \right| $ is the energy of the zero momentum excitation in the superconductor.

We also  argue that an $s$-wave superconductor, in any number of dimensions, is unlikely to have any singularities in its Loschmidt echo. To arrive at these results, we show that the conventional mean field theory used to describe time evolution of out of equilibrium superconductors is not applicable for the purpose of calculating their Loschmidt echo. We give a prescription how this problem can be corrected and show that the time evolution in this problem is equivalent to the Hamiltonian evolution of complex classical spins  satisfying certain boundary conditions. This  construction generalizes   classical Anderson pseudospins (which are real vectors)    used in previous work to describe the time evolution of superconductors. 

In the previous work of some of us \cite{Rylands2021} we claimed that $s$-wave superconductors did have singularities in their echo. We demonstrate here that the quantity evaluated in that work was in fact the {\sl classical echo}, as opposed
to the true quantum echo we calculate below. We also note that Loschmidt echo for topological insulators/superconductors was explored in the literature for quite some time \cite{Dora2015}. However that analysis involved imposing a gap function on a superconductor externally  as opposed to 
determining it self-consistently, and therefore cannot be realized by interacting fermions. 

We begin our analysis with writing down the Hamiltonian for the 2D  chiral $p$-wave superconductor \cite{Gurarie2005,Gurarie2007,Sierra2009,Foster2013}
\be \label{eq:ham1} \hat H = \sum_{\bf p} \xi_p \, \hat a^\dagger_{{\bf p}} \hat a_{{\bf p}} - \frac{g} {V} \sum_{{\bf p}, {\bf q}} p\, q \, e^{i \left( \phi_{\bf p} - \phi_{\bf q} \right)}  \hat a^\dagger_{\bf p}  \hat a^\dagger_{-\bf p}  \hat a_{-{\bf q}}  \hat a_{{\bf q}},
\ee
where  
$p$ 
and $\phi_p$ are the polar coordinates describing the vector ${\bf p}$, 
$\hat a_{\bf p}$ and $\hat a^\dagger_{\bf p}$ are fermionic creation and annihilation operators, $\xi_p = p^2/(2m) - \mu$ with $\mu$ being the chemical potential and $m$ the mass of the fermions, $g$ is the coupling constant and $V$ is the volume of the system. 
It is customary to absorb the angles $\phi_{\bf p}$, $\phi_{\bf q}$ into the fermionic creation and annihilation operators.  Eq.~\rf{eq:ham1} becomes
\be 
\hat H = \sum_{\bf p} \xi_p \, \hat a^\dagger_{{\bf p}} \hat a_{{\bf p}} - \frac{g} {V} \sum_{{\bf p}, {\bf q}} p\, q   \, \hat a^\dagger_{\bf p}  \hat a^\dagger_{-\bf p}  \hat a_{-{\bf q}}  \hat a_{{\bf q}}.
\ee
It is also customary to use mean field theory to study this problem, which results in the simplified
\be \label{eq:hamd} \hat H = \sum_{\bf p} \xi_p \, \hat a^\dagger_{{\bf p}} \hat a_{{\bf p}} - \Delta \sum_{{\bf p}} p   \, \hat a^\dagger_{\bf p}  \hat a^\dagger_{-\bf p}     - \bar \Delta \sum_{\bf p} p \, \hat a_{-{\bf p}}  \hat a_{{\bf p}}.
\ee
Here $\Delta$ and $\bar \Delta$ are generally time-dependent gap functions. These can be related back to the fermions via the so-called gap equation. We will find that the gap equation that we  need here is surprisingly subtle. Because of that, we will postpone its discussion until later and for now  treat $\Delta$ and $\bar \Delta$ as some given time-dependent functions. 

It is well known in the theory of superconductivity that the eigenstates of Eq.~\rf{eq:hamd} can be written in the following form
\be \label{eq:state} \left| \Psi \right> = \prod_{\bf p} \left( u_{\bf p} + v_{\bf p} \, \hat a^\dagger_{\bf p}  \hat a^\dagger_{-\bf p} \right) \left| 0 \right>,
\ee where $\left| 0 \right>$ is a vacuum. Normalization of this wave function requires that  
\be \label{eq:norm} \left| u_{\bf p} \right|^2 + \left| v_{\bf p} \right|^2 = 1. \ee 

In this work we consider the standard quench problem. We initiate  the system in the ground state $\left|\Psi_i \right>$    of the form \rf{eq:state} parametrized by $u_{i {\bf p}}$ and $v_{i {\bf p}}$.
The coupling constant $g$ of the Hamiltonian is suddenly changed to a new value so that $\left| \Psi_i \right>$ is no  longer its eigenstate. The state now evolves forward with the new Hamiltonian. Knowing the time-dependent state
$\left| \Psi(\tau) \right\rangle,$ we can calculate the Loschmidt echo
according to Eq.~\rf{eq:Loschmidt}, or with ${\cal Z} (t) = \left< \Psi_i \Big| \Psi(t) \right\rangle$. 
We would like to understand if there are values of ``Loschmidt time" $t$ at which ${\cal Z}(t)$ is singular. 

We note that evolving the state $\left| \Psi \right>$ forward in time is equivalent to having time dependent $u_{\bf p}(\tau)$ and $v_{\bf p}(\tau)$, satisfying the initial conditions $u_{\bf p}(0)=u_{i \bf p}$, $v_{\bf p}(0)=v_{i \bf p}$. By applying the Hamiltonian
\rf{eq:hamd} to the state \rf{eq:state} we   derive their equations of motion,
\be \label{eq:1} i \dot u_{\bf p} = - \xi_p u_{\bf p} - \bar \Delta \, p \,  v_{\bf p}, \quad i \dot v_{\bf p}  = \xi_p v_{\bf p} - \Delta \, p \, u_{\bf p}.
\ee
To calculate the Loschmidt echo, we evolve $\left| \Psi \right>$ to the Loschmidt time $t$ and project it back onto itself, with the result
\be \label{eq:Z} {\cal Z} = \prod_{\bf p} 2 S_{\bf p}, \quad \ln {\cal Z} = V \int d^2 p \, \ln \left(2 S_{\bf p} \right),
\ee where
\be \label{eq:sp} 2S_{\bf p} = u_{i \bf p}^* u_{\bf p}(t) + v_{i \bf p}^* v_{ \bf p}(t).
\ee
A typical mechanism for ${\cal Z}$ to become singular is for $S_{\bf p}$ to vanish at some critical value $t=t_c$, at some value $p_c$. Let us show that the  $S_0\equiv \lim_{{\bf p} \rightarrow 0} S_{ {\bf p}}$   can vanish in a particularly simple way. Indeed, the equations
of motion of $u_0$, $v_0$
(also understood as limits when ${\bf p}$ is taken to zero) can be solved directly, as they decouple from the functions $\Delta(\tau)$, $\bar \Delta(\tau)$, with the result
\be u_0 = u_{i0} \, e^{i \xi_0 \tau}, \ v_0 = v_{i0} \, e^{-i \xi_0 \tau}.
\ee
Substituting into Eq.~\rf{eq:sp} and taking into account Eq.~\rf{eq:norm} we find
\be 2 S_0 = \cos(\xi_0 t) +  i  \left( u_{i0}^* u_{i0} - v_{i0}^* v_{i0} \right) \sin (\xi_0 t ). \ee
$S_0$ vanishes as a function of $t$   only if \be \label{eq:inplane} \left| u_{i0} \right|^2 = \left|v_{i0} \right|^2. \ee Therefore let us restrict our attention to this case. It is well-known that the $p$-wave superconductor  we study here
can be in the weakly coupled phase or strongly coupled phase depending on the sign of $\xi_0 = -\mu$. The condition \rf{eq:inplane} holds true in the ground state at the critical point between the phases only. Therefore, from now on we consider $\left| \Psi_i \right>$ 
to be the ground state of a critical $p$-wave superconductor, while the Hamiltonian after the quench will describe the strong coupled $\xi_0<0$ or weakly coupled $\xi_0>0$ phase. \ In other words, the chemical potential effectively changes from $\mu_i=0$ to a nonzero
$\mu$ as a result of the quench. It is necessary to keep $\mu$   in the mean-field 
Hamiltonian~(\ref{eq:hamd}) to make sure we have the correct average fermion number, see below and also Ref.~\onlinecite{Rylands2021}. 

With $S_0$ turning to zero at times $t_n = \pi (n+1/2)/\left| \xi_0 \right|$ with integer $n$, ${\cal Z}$ can now be singular at $t=t_n$. However, $S_0$ becoming zero at these times is  not by itself a sufficient condition for ${\cal Z}$ to be singular. To understand if it becomes singular at these times, we need to examine not just the point ${\bf p}=0$ in the momentum space but also its vicinity. Fortunately in this region we can solve the equations of motion \rf{eq:1} perturbatively, using $p$ as a small parameter. 
The solution reads
\begin{eqnarray} u_{\bf p}(\tau) &=& \left( u_{i\bf p} +i v_{i\bf p} \, p \, \int_0^\tau d\tau'  \, \bar \Delta(\tau') \, e^{-2 i \xi_{ p} \tau'} \right)  e^{i \xi_{p} \tau}, \cr v_{\bf p}(\tau) &=&  \left( v_{i\bf p}+i u_{i \bf p} \, p \, \int_0^\tau d\tau'  \,  \Delta(\tau') \, e^{2 i \xi_{p} \tau'} \right)  e^{-i \xi_{p} \tau}. \ \ \ \ \ \ \
\end{eqnarray}
This allows us to calculate, from Eq.~\rf{eq:sp}, 
\begin{eqnarray} \label{eq:sa} 2S_{\bf p} & \approx &  u^*_{i\bf p} u_{i\bf p} \, e^{i \xi_{ p} t} + v^*_{i\bf p} v_{i\bf p} \, e^{-i \xi_{ p} t}  + \cr && i v^*_{i \bf p} u_{i \bf p} \, p \, f_{\bf p} e^{-i \xi_{ p} t} + i u^*_{i \bf p} v_{i \bf p} \, p \,  \bar f_{\bf p} e^{i \xi_{ p} t},
\end{eqnarray}
where
\be f_{\bf p} = \int_0^t d\tau \Delta(\tau) \, e^{2 i \xi_{ p} \tau}, \ \bar f_{\bf p} = \int_0^t d\tau \bar \Delta(\tau) \, e^{-2 i \xi_{ p} \tau}.
\ee
Eq.~\rf{eq:sa} is an expansion in powers of $p$, therefore $u_{i\bf p}$, $v_{i\bf p}$ and $\xi_{p}$ themselves need to be expanded in powers of $p$. Generally, taking into account Eq.~\rf{eq:inplane}, this expansion has the form
\be 2 \left| u_{i \bf p} \right|^2 = 1 +  \alpha   p+ \dots, \ 2 \left| v_{i\bf p} \right|^2 =1 -  \alpha  p + \dots,
\ee where $\alpha$ is some (real) constant. This leads to
\be 2S_p \approx \cos (\xi_0 t) +\oh \left( i \bar f_0 + \alpha \right) p e^{i \xi_0 t} + \oh \left( i f_0 - \alpha \right) p e^{-i \xi_0 t}.
\ee
Let us examine the vicinity of the point in time where $S_0$ vanishes. Expanding in powers of $t-t_n$ and $p$ we find
\be \label{eq:se} 2 S_p \approx (-1)^n \left( \left| \xi_0 \right| \left( t_n-t \right) + \beta p \right), \ \beta = i \alpha + \left( f_0 - \bar f_0 \right)/2.
\ee
We now know, quite generally, the behavior of $S_{\bf p}$ in the vicinity of the point $p=0$ and time $t_n$ where a singularity of ${\cal Z}$ can occur. 
We can estimate the contribution of small ${\bf p}$ to the Loschmidt echo \rf{eq:Z} by writing
\be \frac 1 {\cal Z} \pbyp{\cal Z}{t} = 2\pi V \int_0^{p_0}  \frac{p \, dp}{ t-t_n  - \beta p/\left| \xi_0 \right|},
\ee 
where $p_0$ is some momentum beyond which the expansion \rf{eq:se} no longer holds. The integral is easy to evaluate and produces
\be \label{eq:qu} \frac 1 {V \cal Z}  \pbyp{\cal Z}{t} = \frac{2 \pi \xi_0^2 \left( t - t_n \right)  }{\beta^2  }  \ln \left[ \frac { \left| \xi_0 \right|  \left( t - t_n \right)} {\beta   p_0} \right]
\ee
as the singular contribution to the Loschmidt echo (with the second derivative of $\ln {\cal Z}$ over time $t$ having logarithmic singularity). 

Eq.~\rf{eq:qu} constitutes the main result of this work. A 2D $p$-wave superconductor when quenched out of its critical point into either its weak (topological) or strong (non-topological) phase has periodic singularities 
in its Loschmidt echo where its second derivative over time diverges logarithmically. The singularity is driven by the behavior of small momentum fermions. The parameter $\alpha$ which appears in the calculations above is controlled 
by the initial amplitudes $u_{i \bf p}$, $v_{i \bf p}$. Those can be found as the ground state of the Hamiltonian before quench is known. Being at the critical point where $\xi_{i0}=-\mu_i=0$, it must have the form
\be \label{eq:ham0} \hat H_{i} = \sum_{\bf p} \frac{p^2}{2m} \, \hat a^\dagger_{{\bf p}} \hat a_{{\bf p}} - \Delta_i \sum_{{\bf p}} p   \, \hat a^\dagger_{\bf p}  \hat a^\dagger_{-\bf p}     - \bar \Delta_i \sum_{\bf p} p \, \hat a_{-{\bf p}}  \hat a_{{\bf p}}.
\ee
where $\Delta_i$ and $\bar \Delta_i$ are the equilibrium gap functions. Calculating its ground state is a standard exercise. Evaluating $u_{i \bf p}$ and $v_{i \bf p}$ leads to
$ \alpha = 1 /\left( 4 m\sqrt{\bar \Delta_i \Delta_i} \right)$.

We would like to point out that the ratio of the amplitudes $v_{i \bf p}/u_{i \bf p}$  goes either to infinity or to zero in the two phases of the 2D chiral $p$-wave superconductors as $p\to0$, determining the topological properties \cite{Read2000}
of these phases. The only exception is the critical point between the phases where this ratio is $1$. Therefore, the criticality in the Loschmidt echo directly  reflects the fact that before the quench the system is at the critical point between the two phases with
different topology.

The question still remains whether $S_{\bf p}$ can also vanish at other values of momenta, perhaps leading to other singularities in ${\cal Z}$ which we also need to explore. We would like to argue that this does not happen. To do this, we
need to understand further how $\Delta$ and $\bar \Delta$ are related to the fermions. Usually in the problem of quantum quench where the goal is to calculate $\left| \Psi(\tau) \right>$ after the quench, the following equation is used for $\Delta$,
\be \label{eq:oldsaddle} \Delta(\tau) = \frac{g}{V} \sum_{\bf p} p\, \left< \Psi(\tau) \right| \hat a_{-{\bf p}}  \hat a_{{\bf p}} \left| \Psi(\tau) \right> =\frac g V \sum_{\bf p} p \, u^*_{\bf p}(\tau) v_{\bf p}(\tau), 
\ee
and its complex conjugate for $\bar \Delta(\tau)$. Substituting this into Eq.~\rf{eq:1} produces the equations of motion for $u_{\bf p}$ and $v_{\bf p}$. They are nonlinear but known to be integrable. Their solution can be found for a variety of initial conditions 
and allows to calculate $\left| \Psi(t) \right>$ in many interesting cases~\cite{Foster2013}. 

However, we would like to argue that these equations are not suitable for calculating the Loschmidt echo \rf{eq:Loschmidt}. The resulting wave function $\left| \Psi(t) \right\rangle$, while  well suited for calculating the expectations of local observables 
in the problem and finding their time-dependence, produces wrong results if  used to evaluate  overlaps of $\left| \Psi_i \right\rangle$ and $\left| \Psi(t) \right\rangle$ occurring in the calculation of ${\cal Z}$. 

While leaving the detailed construction for the Supplemental Material, we note that the Loschmidt echo is simply a matrix element of the evolution operator. These can be calculated using conventional Feynman functional integral,
avoiding the intricacies involved in the Schwinger-Keldysh functional integral construction. The saddle point approximation with respect to $\Delta$ and $\bar \Delta$ calculated in the 
framework of the conventional Feynman functional integral produces (see also Ref.~\cite{Eckstein2014})
\be \label{eq:newgap} \Delta(\tau) = \frac g {V {\cal Z}} \sum_{\bf p} p \left\langle \Psi_i \right| e^{-i \hat H (t-\tau)}  \hat a_{-{\bf p}}  \hat a_{{\bf p}}  e^{-i \hat H \tau} \left| \Psi_i \right\rangle,
\ee
and similarly for $\bar \Delta(\tau)$. This equation replaces the equation \rf{eq:oldsaddle} for the purpose of determining $\Delta$ and $\bar \Delta$. 

To elucidate the meaning of Eq.~\rf{eq:newgap} we introduce a new wave function defined by
\be \left| \tilde \Psi(\tau) \right\rangle = e^{i\hat H (t-\tau)} \left| \Psi_i \right\rangle.
\ee
Just like $\left| \Psi(\tau) \right>$, it is described by the amplitudes $\tilde u_{\bf p}(\tau)$, $\tilde v_{\bf p}(\tau)$. They satisfy the same equations of motion \rf{eq:1} as $u_{\bf p}(\tau)$, $v_{\bf p}(\tau)$, but with different boundary conditions. Whereas $u_{\bf p}(0)= u_{i \bf p}$, $v_{\bf p}(0)=v_{i \bf p}$, the boundary conditions on these new amplitudes are imposed at $t=\tau$, $\tilde u_{\bf p}(t)= u_{i \bf p}$, $\tilde v_{\bf p}(t)=v_{i \bf p}$. In terms of these we find
\be \frac{1}{\cal Z}  \left\langle \Psi_i \right| e^{-i \hat H (t-\tau)}  \hat a_{-{\bf p}}  \hat a_{{\bf p}}  e^{-i \hat H \tau} \left| \Psi_i \right>  =\frac{ \tilde u^*_{\bf p}(\tau) v_{\bf p}(\tau)}{2 S_{\bf p}}, 
\ee
where  $S_{\bf p}$ can also be expressed in terms of these amplitudes by
\be \label{eq:spinlength} 2 S_{\bf p}(t) = \tilde u^*_{\bf p} (\tau) u_{\bf p}(\tau) + \tilde v^*_{\bf p} (\tau) v_{\bf p}(\tau).
\ee
Note that $S_{\bf p}$ do not depend on $\tau$,  thus they can be calculated at arbitrary $\tau$. In particular, substituting $\tau=t$ we see that the definitions \rf{eq:spinlength} and  \rf{eq:sp} coincide. With the help of these relations we find
\be \label{eq:newgap1} \Delta (\tau) = \frac g V \sum_{\bf p} \frac{p \, \tilde u^*_{\bf p}(\tau) v_{\bf p}(\tau)}{2S_{\bf p}}, \ \bar \Delta(\tau) = \frac g V \sum_{\bf p} \frac{p \, \tilde v^*_{\bf p}(\tau) u_{\bf p}(\tau)}{2S_{\bf p}}.
\ee
These new gap equations replace the old equation \rf{eq:oldsaddle}. Thus we now need to solve equations of motion \rf{eq:1} supplemented by Eq.~\rf{eq:newgap1}. These equations constitute the second main result of this work. Note that unlike in
Eq.~\rf{eq:oldsaddle}, here $\bar \Delta$ is not equal to the complex conjugate of $\Delta$. 

Solving these new equations of motion is an interesting problem by itself,  which will be left as a subject for future work. 
Here  we would just like to see whether they are compatible with the singularities in ${\cal Z}$ that we found earlier. In the quench scenario considered earlier, we had $S_{\bf p}$ vanish for small ${\bf p}$ 
according to Eq.~\rf{eq:se}. Substituting this into Eqs.~\rf{eq:newgap1} we find
\be \Delta \sim \int \frac{\tilde u^*_{\bf p}(\tau) v_{\bf p}(\tau)\,  p^2 dp} {\left| \xi_0 \right| \left(t_n - t \right) + \beta p}.
\ee
This shows that as a function of $t$, $\Delta$ will have a divergent second derivative. This by itself does not affect the earlier established fact of the divergent second derivative of ${\cal Z}$. 

However, up until now we only analyzed vanishing of $S_0$. What if $S_{\bf p}$ vanishes for some nonzero $p_c$? If this were to happen, then the gap equation should be expected to read
\be \Delta \sim \int \frac{\tilde u^*_{\bf p}(\tau) v_{\bf p}(\tau) \, p^2 dp} {t-t_c + \beta (p-p_c)},
\ee
where the expression to be integrated is an approximation valid in the vicinity of $p \sim p_c$. 
Taking into account that $\beta$ is generally complex and recalling the standard formula
${\rm Im} \, 1/(x \pm i \epsilon) = \mp i \pi \delta(x),$
we see that $\Delta$ will then generally be a discontinuous function of $t$. If $\Delta$ is discontinuous, so will be ${\cal Z}$. On the other hand, ${\cal Z}$ is closely related to the partition functions of quantum systems. Partition functions of thermal systems cannot be discontinuous functions of temperature. Likewise we expect that the Loschmidt echo cannot be a discontinuous function of time $t$, thus we conclude that the scenario where $S_{\bf p}$ vanishes for some nonzero ${p}$  cannot be realized. 

Note that the $s$-wave superconductors are described by the Hamiltonians very similar to the ones studied here, but without the extra factors of momenta in the interactions. Let us test if they could have a vanishing $S_{\bf p}$ at some
critical $p_c$. 
Putting this superconductor in $d$ dimensional space for generality, the gap equation and the expression for the Loschmidt echo now read
$$
 \Delta \sim \int \frac{\tilde u^*_{\bf p}(\tau) v_{\bf p}(\tau) p^{d-1} dp} {t-t_c + \beta (p-p_c)}, \ \frac{\partial {\cal Z}}{\partial t} \sim \int \frac{p^{d-1}dp} { t-t_c + \beta (p-p_c)}.
$$
Again, unless $p_c=0$, the equations above lead to the discontinuity of $\Delta$  as a function of $t$ and therefore the discontinuity in ${\cal Z}$, which we expect cannot happen. The only exception would be  $p_c=0$. 
However,  here  ${\bf p}=0$ is not special in the same way as in $p$-wave superconductors as the spin at ${\bf p}=0$ is coupled to the rest of the spins and we were unable to identify any initial conditions for which $S_0$ can vanish.
We are also not aware of any alternative 
calculation showing vanishing of $S_0$ in $s$-wave superconductors. 


On the contrary, we expect that the criticality in Loschmidt echo due to the ${\bf p}=0$ mode also manifests itself in the higher-order superconductors, such as 2D $d_{x^2-y^2}+i d_{xy}$ superconductor. Exploring this could be the subject of further research.

Given the wave function $\left| \Psi(t) \right\rangle$ calculated using the standard approach of Eqs.~\rf{eq:1} together with \rf{eq:oldsaddle}, one can ask whether its overlap with the initial wave function $\left|\Psi_i\right\rangle$ is still meaningful. 
Let us argue that 
\be \label{eq:rylands} {\cal L} = \left| \left< \Psi_i \Big| \Psi(t) \right> \right|^2 
\ee coincides with the classical echo defined as \cite{Prosen2004}
\be \label{eq:prosen} {\cal L} = \int d{\bf x} \, \rho({\bf x}, 0) \rho({\bf x},  t).
\ee
Here ${\bf x}$  are the coordinates parametrizing the phase space of a classical system and $\rho({\bf x}, t)$ is the classical distribution function, which is in general time-dependent. 

Indeed, equations of evolution of $\left| \Psi(t) \right\rangle$ \rf{eq:1} together with \rf{eq:oldsaddle} are quasiclassical and should be equivalent to evolving the classical distribution function $\rho$. Therefore, it should not be surprising that
Eqs.~\rf{eq:rylands} and \rf{eq:prosen} coincide. Formal proof of that consists of identifying $\rho$ for the interacting fermions system that we study here with the Wigner function computed from the quantum state of our system and 
showing formally that Eq.~\rf{eq:rylands} reduces to \rf{eq:prosen}; see Supplemental Material to see how this calculation can be carried out. 
The quantity ${\cal L}$ was calculated for $s$- and $p$-wave superconductors in Ref.~\onlinecite{Rylands2021} and  found to have many singularities as a function of $t$. Thus we arrive at a striking conclusion: the classical echo \rf{eq:prosen} can be singular, even when
the full quantum echo calculated here is not.

An interesting remaining question is the role of the chemical potential $\mu$  we introduced into the post-quench Hamiltonian. Normally in dynamical problems with conserved total particle number $\mu$ is arbitrary as changing $\mu$ simply changes the phase of the time-dependent wave function. However, our initial wave function is not an eigenstate of  the total particle number operator $\hat N = \sum_{\bf p} \hat a^\dagger_{\bf p} \hat a_{\bf p}$. 
 Note that $\left\langle \Psi(\tau) \right| \hat N \left| \Psi(\tau) \right\rangle$ can be expressed entirely in terms of $u_{i \bf p}$, $v_{i \bf p}$ characterizing the initial wave function which  contains $\mu_i$ but is independent of $\mu$. A more relevant expectation value
which arises in the process of evaluating the Loschmidt echo and depends on $\mu$ is
$\left\langle \Psi_i \right| e^{-i \hat H (t-\tau)}
\hat N e^{-i \hat H \tau} \left| \Psi_i \right\rangle/{\cal Z}$. Therefore, $\mu$ must be chosen in such a way as to make this expectation equal to the desired fermion number, i.e., we need to introduce $\mu$ to ensure that we describe the time evolution with the correct number of fermions. More details of this
formalism is given in the Supplemental material.

{\sl Acknowledgement:} We are grateful to C. Rylands and V. Galitski for useful conversations at the beginning of this project.  VG was supported by the Simons Collaboration on Ultra-Quantum Matter, which is a grant from the Simons Foundation (651440).

\bibliography{library}

\begin{thebibliography}{12}%
\makeatletter
\providecommand \@ifxundefined [1]{%
 \@ifx{#1\undefined}
}%
\providecommand \@ifnum [1]{%
 \ifnum #1\expandafter \@firstoftwo
 \else \expandafter \@secondoftwo
 \fi
}%
\providecommand \@ifx [1]{%
 \ifx #1\expandafter \@firstoftwo
 \else \expandafter \@secondoftwo
 \fi
}%
\providecommand \natexlab [1]{#1}%
\providecommand \enquote  [1]{``#1''}%
\providecommand \bibnamefont  [1]{#1}%
\providecommand \bibfnamefont [1]{#1}%
\providecommand \citenamefont [1]{#1}%
\providecommand \href@noop [0]{\@secondoftwo}%
\providecommand \href [0]{\begingroup \@sanitize@url \@href}%
\providecommand \@href[1]{\@@startlink{#1}\@@href}%
\providecommand \@@href[1]{\endgroup#1\@@endlink}%
\providecommand \@sanitize@url [0]{\catcode `\\12\catcode `\$12\catcode
  `\&12\catcode `\#12\catcode `\^12\catcode `\_12\catcode `\%12\relax}%
\providecommand \@@startlink[1]{}%
\providecommand \@@endlink[0]{}%
\providecommand \url  [0]{\begingroup\@sanitize@url \@url }%
\providecommand \@url [1]{\endgroup\@href {#1}{\urlprefix }}%
\providecommand \urlprefix  [0]{URL }%
\providecommand \Eprint [0]{\href }%
\providecommand \doibase [0]{http://dx.doi.org/}%
\providecommand \selectlanguage [0]{\@gobble}%
\providecommand \bibinfo  [0]{\@secondoftwo}%
\providecommand \bibfield  [0]{\@secondoftwo}%
\providecommand \translation [1]{[#1]}%
\providecommand \BibitemOpen [0]{}%
\providecommand \bibitemStop [0]{}%
\providecommand \bibitemNoStop [0]{.\EOS\space}%
\providecommand \EOS [0]{\spacefactor3000\relax}%
\providecommand \BibitemShut  [1]{\csname bibitem#1\endcsname}%
\let\auto@bib@innerbib\@empty
\bibitem [{\citenamefont {Heyl}\ \emph {et~al.}(2013)\citenamefont {Heyl},
  \citenamefont {Polkovnikov},\ and\ \citenamefont {Kehrein}}]{Kehrein2013}%
  \BibitemOpen
  \bibfield  {author} {\bibinfo {author} {\bibfnamefont {M.}~\bibnamefont
  {Heyl}}, \bibinfo {author} {\bibfnamefont {A.}~\bibnamefont {Polkovnikov}}, \
  and\ \bibinfo {author} {\bibfnamefont {S.}~\bibnamefont {Kehrein}},\ }\href
  {\doibase 10.1103/PhysRevLett.110.135704} {\bibfield  {journal} {\bibinfo
  {journal} {Phys. Rev. Lett.}\ }\textbf {\bibinfo {volume} {110}},\ \bibinfo
  {pages} {135704} (\bibinfo {year} {2013})}\BibitemShut {NoStop}%
\bibitem [{Note1()}]{Note1}%
  \BibitemOpen
  \bibinfo {note} {The standard definition of the Loschdmidt echo is $\left |
  \left < \Psi _i \right | e^{i H_2 t} e^{-i H_1 t} \left | \Psi \right >
  \right |^2$. If $\Psi $ is the eigenstate of $H_1$, up to taking its absolute
  value square it effectively reduces to Eq.~(\ref
  {eq:Loschmidt}).}\BibitemShut {Stop}%
\bibitem [{\citenamefont {Read}\ and\ \citenamefont {Green}(2000)}]{Read2000}%
  \BibitemOpen
  \bibfield  {author} {\bibinfo {author} {\bibfnamefont {N.}~\bibnamefont
  {Read}}\ and\ \bibinfo {author} {\bibfnamefont {D.}~\bibnamefont {Green}},\
  }\href {\doibase 10.1103/PhysRevB.61.10267} {\bibfield  {journal} {\bibinfo
  {journal} {Phys. Rev. B}\ }\textbf {\bibinfo {volume} {61}},\ \bibinfo
  {pages} {10267} (\bibinfo {year} {2000})}\BibitemShut {NoStop}%
\bibitem [{\citenamefont {C.~Rylands}\ \emph {et~al.}(2021)\citenamefont
  {C.~Rylands}, \citenamefont {Yuzbashyan}, \citenamefont {Gurarie},
  \citenamefont {Zabalo},\ and\ \citenamefont {Galitski}}]{Rylands2021}%
  \BibitemOpen
  \bibfield  {author} {\bibinfo {author} {\bibfnamefont {C.}~\bibnamefont
  {C.~Rylands}}, \bibinfo {author} {\bibfnamefont {E.~A.}\ \bibnamefont
  {Yuzbashyan}}, \bibinfo {author} {\bibfnamefont {V.}~\bibnamefont {Gurarie}},
  \bibinfo {author} {\bibfnamefont {A.}~\bibnamefont {Zabalo}}, \ and\ \bibinfo
  {author} {\bibfnamefont {V.}~\bibnamefont {Galitski}},\ }\href {\doibase
  https://doi.org/10.1016/j.aop.2021.168554} {\bibfield  {journal} {\bibinfo
  {journal} {Ann. Phys.}\ }\textbf {\bibinfo {volume} {435}},\ \bibinfo {pages}
  {168554} (\bibinfo {year} {2021})}\BibitemShut {NoStop}%
\bibitem [{\citenamefont {Vajna}\ and\ \citenamefont
  {D\'ora}(2015)}]{Dora2015}%
  \BibitemOpen
  \bibfield  {author} {\bibinfo {author} {\bibfnamefont {S.}~\bibnamefont
  {Vajna}}\ and\ \bibinfo {author} {\bibfnamefont {B.}~\bibnamefont {D\'ora}},\
  }\href {\doibase 10.1103/PhysRevB.91.155127} {\bibfield  {journal} {\bibinfo
  {journal} {Phys. Rev. B}\ }\textbf {\bibinfo {volume} {91}},\ \bibinfo
  {pages} {155127} (\bibinfo {year} {2015})}\BibitemShut {NoStop}%
\bibitem [{\citenamefont {Gurarie}\ \emph {et~al.}(2005)\citenamefont
  {Gurarie}, \citenamefont {Radzihovsky},\ and\ \citenamefont
  {Andreev}}]{Gurarie2005}%
  \BibitemOpen
  \bibfield  {author} {\bibinfo {author} {\bibfnamefont {V.}~\bibnamefont
  {Gurarie}}, \bibinfo {author} {\bibfnamefont {L.}~\bibnamefont
  {Radzihovsky}}, \ and\ \bibinfo {author} {\bibfnamefont {A.~V.}\ \bibnamefont
  {Andreev}},\ }\href {\doibase 10.1103/PhysRevLett.94.230403} {\bibfield
  {journal} {\bibinfo  {journal} {Phys. Rev. Lett.}\ }\textbf {\bibinfo
  {volume} {94}},\ \bibinfo {pages} {230403} (\bibinfo {year}
  {2005})}\BibitemShut {NoStop}%
\bibitem [{\citenamefont {Gurarie}\ and\ \citenamefont
  {Radzihovsky}(2007)}]{Gurarie2007}%
  \BibitemOpen
  \bibfield  {author} {\bibinfo {author} {\bibfnamefont {V.}~\bibnamefont
  {Gurarie}}\ and\ \bibinfo {author} {\bibfnamefont {L.}~\bibnamefont
  {Radzihovsky}},\ }\href {\doibase https://doi.org/10.1016/j.aop.2006.10.009}
  {\bibfield  {journal} {\bibinfo  {journal} {Ann. Phys.}\ }\textbf {\bibinfo
  {volume} {322}},\ \bibinfo {pages} {2} (\bibinfo {year} {2007})}\BibitemShut
  {NoStop}%
\bibitem [{\citenamefont {Iba\~nez}\ \emph {et~al.}(2009)\citenamefont
  {Iba\~nez}, \citenamefont {Links}, \citenamefont {Sierra},\ and\
  \citenamefont {Zhao}}]{Sierra2009}%
  \BibitemOpen
  \bibfield  {author} {\bibinfo {author} {\bibfnamefont {M.}~\bibnamefont
  {Iba\~nez}}, \bibinfo {author} {\bibfnamefont {J.}~\bibnamefont {Links}},
  \bibinfo {author} {\bibfnamefont {G.}~\bibnamefont {Sierra}}, \ and\ \bibinfo
  {author} {\bibfnamefont {S.-Y.}\ \bibnamefont {Zhao}},\ }\href {\doibase
  10.1103/PhysRevB.79.180501} {\bibfield  {journal} {\bibinfo  {journal} {Phys.
  Rev. B}\ }\textbf {\bibinfo {volume} {79}},\ \bibinfo {pages} {180501}
  (\bibinfo {year} {2009})}\BibitemShut {NoStop}%
\bibitem [{\citenamefont {Foster}\ \emph {et~al.}(2013)\citenamefont {Foster},
  \citenamefont {Dzero}, \citenamefont {Gurarie},\ and\ \citenamefont
  {Yuzbashyan}}]{Foster2013}%
  \BibitemOpen
  \bibfield  {author} {\bibinfo {author} {\bibfnamefont {M.~S.}\ \bibnamefont
  {Foster}}, \bibinfo {author} {\bibfnamefont {M.}~\bibnamefont {Dzero}},
  \bibinfo {author} {\bibfnamefont {V.}~\bibnamefont {Gurarie}}, \ and\
  \bibinfo {author} {\bibfnamefont {E.~A.}\ \bibnamefont {Yuzbashyan}},\ }\href
  {\doibase 10.1103/PhysRevB.88.104511} {\bibfield  {journal} {\bibinfo
  {journal} {Phys. Rev. B}\ }\textbf {\bibinfo {volume} {88}},\ \bibinfo
  {pages} {104511} (\bibinfo {year} {2013})}\BibitemShut {NoStop}%
\bibitem [{\citenamefont {Canovi}\ \emph {et~al.}(2014)\citenamefont {Canovi},
  \citenamefont {Werner},\ and\ \citenamefont {Eckstein}}]{Eckstein2014}%
  \BibitemOpen
  \bibfield  {author} {\bibinfo {author} {\bibfnamefont {E.}~\bibnamefont
  {Canovi}}, \bibinfo {author} {\bibfnamefont {P.}~\bibnamefont {Werner}}, \
  and\ \bibinfo {author} {\bibfnamefont {M.}~\bibnamefont {Eckstein}},\ }\href
  {\doibase 10.1103/PhysRevLett.113.265702} {\bibfield  {journal} {\bibinfo
  {journal} {Phys. Rev. Lett.}\ }\textbf {\bibinfo {volume} {113}},\ \bibinfo
  {pages} {265702} (\bibinfo {year} {2014})}\BibitemShut {NoStop}%
\bibitem [{\citenamefont {Veble}\ and\ \citenamefont
  {Prosen}(2004)}]{Prosen2004}%
  \BibitemOpen
  \bibfield  {author} {\bibinfo {author} {\bibfnamefont {G.}~\bibnamefont
  {Veble}}\ and\ \bibinfo {author} {\bibfnamefont {T.}~\bibnamefont {Prosen}},\
  }\href {\doibase 10.1103/PhysRevLett.92.034101} {\bibfield  {journal}
  {\bibinfo  {journal} {Phys. Rev. Lett.}\ }\textbf {\bibinfo {volume} {92}},\
  \bibinfo {pages} {034101} (\bibinfo {year} {2004})}\BibitemShut {NoStop}%
\bibitem [{\citenamefont {Davis}\ \emph {et~al.}(2021)\citenamefont {Davis},
  \citenamefont {Kumari}, \citenamefont {Mann},\ and\ \citenamefont
  {Ghose}}]{Davis2021}%
  \BibitemOpen
  \bibfield  {author} {\bibinfo {author} {\bibfnamefont {J.}~\bibnamefont
  {Davis}}, \bibinfo {author} {\bibfnamefont {M.}~\bibnamefont {Kumari}},
  \bibinfo {author} {\bibfnamefont {R.~B.}\ \bibnamefont {Mann}}, \ and\
  \bibinfo {author} {\bibfnamefont {S.}~\bibnamefont {Ghose}},\ }\href
  {\doibase 10.1103/PhysRevResearch.3.033134} {\bibfield  {journal} {\bibinfo
  {journal} {Phys. Rev. Research}\ }\textbf {\bibinfo {volume} {3}},\ \bibinfo
  {pages} {033134} (\bibinfo {year} {2021})}\BibitemShut {NoStop}%
\end{thebibliography}%

\renewcommand{\bibnumfmt}[1]{[S#1]}
\renewcommand{\theequation}{S\arabic{equation}}
\renewcommand{\thefigure}{S\arabic{figure}}
\renewcommand{\thetable}{S.\Roman{table}}

\setcounter{equation}{0}
\setcounter{figure}{0}
\setcounter{table}{0}

\clearpage
\onecolumngrid


\section{Supplemental Material}
\onecolumngrid

\subsection{A: Hamiltonian for the $p$-wave interacting fermions}
Spinless or spin polarized fermions cannot interact in the $s$-wave channel. Their strongest interactions are in the $p$-wave channel. In 2D space, their Hamiltonian takes the following form
\be \label{eq:ham1s} \hat H = \sum_{\bf p} \xi_p \, \hat a^\dagger_{{\bf p}} \hat a_{{\bf p}} - \frac{2g} {V} \sum_{{\bf p}, {\bf q}, {\bf k}} {\bf p} \cdot {\bf q} \,  \, \hat a^\dagger_{\frac{\bf k}{2} +\bf p}  \hat a^\dagger_{\frac{\bf k}{2} -\bf p}  \hat a_{\frac{\bf k}{2} -{\bf q}}  \hat a_{\frac{\bf k}{2} +{\bf q}},
\ee
Here \be \xi_p=\frac{p^2}{2m} - \mu, \ee $\hat a^\dagger_{\bf p}$ and $\hat a_{\bf p}$ are the creation and annihilation operators for the spinless fermions, $g$ is the coupling constant and $V$ is the volume of the system. 
In the applications to superconductivity, only the terms with ${\bf k}=0$ play a substantial role. From now on we will therefore remove summation over ${\bf k}$ and set ${\bf k}=0$ with the result
\be \label{eq:ham2s} \hat H = \sum_{\bf p} \xi_p \, \hat a^\dagger_{{\bf p}} \hat a_{{\bf p}} - \frac{2g} {V} \sum_{{\bf p}, {\bf q}} {\bf p} \cdot {\bf q} \,  \, \hat a^\dagger_{\bf p}  \hat a^\dagger_{-\bf p}  \hat a_{-{\bf q}}  \hat a_{{\bf q}},
\ee
The most interesting superconducting phase described by this Hamiltonian is the chiral phase~\cite{Gurarie2005} where 
\be \label{eq:chirals} \left\langle \hat a_{-{\bf p}}  \hat a_{{\bf p}} \right\rangle \sim e^{i \phi_{\bf p}},\ee
where $\phi_{\bf p}$ is the polar angle of the vector ${\bf p}$ in the 2D space. It is also often referred to as $p_x + i p_y$ phase of the superconductor. In this phase, it is advantageous to split the interaction according to
\be \label{eq:splits} {\bf p} \cdot {\bf q} = \oh \left[ \left(p_x + i p_y \right) \left( q_x - i q_y \right) + \left(p_x - i p_y \right) \left( q_x + i q_y \right) \right]. 
\ee
The right-most term in \rf{eq:splits}, when substituted into Eq.~\rf{eq:ham2s}, produces zero in the $p_x+ip_y$ phase and therefore can be removed altogether, with the result
\be \label{eq:ham3s} \hat H = \sum_{\bf p} \xi_p \, \hat a^\dagger_{{\bf p}} \hat a_{{\bf p}} - \frac{g} {V} \sum_{{\bf p}, {\bf q}} p\, q \, e^{i \left( \phi_{\bf p} - \phi_{\bf q} \right)}  \hat a^\dagger_{\bf p}  \hat a^\dagger_{-\bf p}  \hat a_{-{\bf q}}  \hat a_{{\bf q}},
\ee
Importantly, this nullification happens in any state, ground state or excited state, where Eq.~\rf{eq:chirals} holds. Here we restrict ourselves only to such states. In particular, we are going to consider time evolution of some initial state. All such
initial states will be chosen to satisfy Eq.~\rf{eq:chirals}. Then we can fully rely on Eq.~\rf{eq:ham3s} as the Hamiltonian describing the time evolution.

The final step consists of absorbing the phases $\phi_{\bf p}$ and $\phi_{\bf q}$ in the creation and annihilation operators 
\be e^{i \phi_{\bf p}/2} \hat a^\dagger_{\bf p}  \rightarrow \hat a^\dagger_{\bf p}, \ e^{-i  \phi_{\bf p}/ 2 } \hat a_{\bf p}  \rightarrow \hat a_{\bf p}, \ee
with the result
\be  \label{eq:ham4s} \hat H = \sum_{\bf p} \xi_p \, \hat a^\dagger_{{\bf p}} \hat a_{{\bf p}} - \frac{g} {V} \sum_{{\bf p}, {\bf q}} p\, q \,   \hat a^\dagger_{\bf p}  \hat a^\dagger_{-\bf p}  \hat a_{-{\bf q}}  \hat a_{{\bf q}}.
\ee
This is the Hamiltonian that we will use throughout this work.

\subsection{B: Saddle point approximation for Loschmidt echo}
We are interested in calculating the Loschmidt echo, that is the quantity
\be {\cal Z} = \left\langle \Psi_i \right| e^{-i \hat H t} \left| \Psi_i \right\rangle.
\ee
Here $\left| \Psi_i \right>$ is some state which is not an eigenstate of $\hat H$. We would like to use mean field theory to do this calculation, as common in the theory of superconductivity. One can argue that as long as the total number of
fermions is large, mean field theory is a good approximation in the Hamiltonians \rf{eq:ham2s}, \rf{eq:ham3s}, \rf{eq:ham4s} (on the contrary, in the Hamiltonian \rf{eq:ham1s} mean field theory can sometimes fail, therefore in passing from Eq.~\rf{eq:ham1s} to \rf{eq:ham2s} we 
implicitly assumed that mean theory works). 

Setting up mean field theory for purpose of calculating Loschmidt echo is subtle. Let us set up the appropriate formalism to do it. 

We express the echo in terms of the coherent state functional integral
\be {\cal Z} = \int {\cal D} \psi {\cal D} \bar \psi \, e^{i \int_0^t d\tau \left\{ \sum_{\bf p} \left[ \bar \psi_{\bf p} \left( i \pp{\tau} - \xi_{p} \right) \psi_{\bf p} \right] + \frac g V \sum_{{\bf p}, {\bf q}} p\, q \,   \bar \psi_{\bf p}  \bar \psi_{-\bf p}  \psi_{-{\bf q}}  \psi_{{\bf q}}
 \right\} }.
\ee 
In order to represent the required matrix element of the evolution operator, the fields $\psi_{\bf p}$ and $\bar \psi_{\bf p}$ must satisfy the appropriate boundary conditions at $\tau=0$ and $\tau=t$ whose specific form will not be important here.

As common in the theory of superconductivity we introduce the Hubbard-Stratonovich field $\Delta$, which results in
\be {\cal Z} =\int {\cal D} \Delta {\cal D} \bar \Delta  \, e^{i W},
\ee where
\begin{eqnarray} \label{eq:HSs} e^{i W} &=& \int {\cal D} \psi {\cal D} \bar \psi   \, e^{i S}, \cr S &=& i \int_0^t d\tau \left\{ \sum_{\bf p} \left[ \bar \psi_{\bf p} \left( i \pp{\tau} - \xi_{p} \right) \psi_{\bf p} \right] + \Delta \sum_{{\bf p}} p \,  \bar \psi_{\bf p} \bar \psi_{-\bf p} + \bar \Delta  \sum_{{\bf p}} p   \, \psi_{-{\bf p}} \psi_{\bf p} - \frac{V}{g} \bar \Delta
\Delta  \right\}.
\end{eqnarray}
We calculate the integral over $\Delta$ and $\bar \Delta$ in the saddle point approximation. Varying $W$ over $\bar \Delta(\tau)$ at some time $\tau$, we find
\be \frac 1 {\cal Z} \int {\cal D} \psi {\cal D} \bar \psi   \, \left( \sum_{{\bf p}} p   \, \psi_{-{\bf p}} (\tau)  \psi_{\bf p} (\tau) - \frac V g \Delta(\tau)  \right) e^{iS}=0.
\ee 
Finally we reinterpret the resulting functional integral as 
\be \label{eq:gaps} \Delta(\tau) =\frac g {V {\cal Z}} \sum_{\bf p} p \left\langle \Psi_i \right| e^{-i \hat H (t- \tau) } \hat a_{-\bf p} \hat a_{\bf p} e^{-i \hat H \tau} \left| \Psi_i \right\rangle.
\ee Similarly, by varying $W$ over $\Delta$ we can find
\be \label{eq:gapcs} \bar \Delta(\tau) =\frac g {V\cal Z} \sum_{\bf p} p \left\langle \Psi_i \right| e^{-i \hat H (t- \tau) } \hat a^\dagger_{\bf p} \hat a^\dagger_{-\bf p} e^{-i \hat H \tau} \left| \Psi_i \right\rangle.
\ee
This is the version of the saddle point (or gap) equation appropriate for calculating the Loschmidt echo, which is used in the main text. Interestingly, Eq.~\rf{eq:gapcs} is not the complex conjugate of Eq.~\rf{eq:gaps},
unlike in the more common occurrences of the gap equation where usually $\Delta$ and $\bar \Delta$ are complex conjugates of each other. 

The mean field Hamiltonian follows from Eq.~\rf{eq:HSs} to be
\be \label{eq:ham5s} \hat H = \sum_{\bf p} \xi_p \, \hat a^\dagger_{{\bf p}} \hat a_{{\bf p}} - \Delta \sum_{{\bf p}} p   \, \hat a^\dagger_{\bf p}  \hat a^\dagger_{-\bf p}     - \bar \Delta \sum_{\bf p} p \, \hat a_{-{\bf p}}  \hat a_{{\bf p}}.
\ee

The initial state $\left| \Psi_i \right>$ in our formalism is a coherent state and so does not have a fixed number of particles. To fix the particle number we can introduce the time dependent chemical potential as a Lagrange multiplier. That generalizes the 
formalism to include
\be {\cal Z} =\int {\cal D} \Delta {\cal D} \bar \Delta  {\cal D} \mu \, e^{i W -i N \int_0^t d\tau \, \mu(\tau)},
\ee where
\be  \label{eq:HS1s} e^{i W} = \int {\cal D} \psi {\cal D} \bar \psi   \, e^{i S[\mu]}, \ee
 $$S[\mu] = i \int_0^t d\tau \left\{ \sum_{\bf p} \left[ \bar \psi_{\bf p} \left( i \pp{\tau} - \frac{p^2}{2m} +\mu(\tau) \right) \psi_{\bf p} \right] + \Delta \sum_{{\bf p}} p \,  \bar \psi_{\bf p} \bar \psi_{-\bf p} + \bar \Delta  \sum_{{\bf p}} p   \, \psi_{-{\bf p}} \psi_{\bf p} - \frac{V}{g} \bar \Delta
\Delta  \right\}. $$
We now apply the saddle point approximation not only over $\Delta$ and $\bar \Delta$, but also over $\mu$. The saddle point over $\mu$ gives
\be N = \frac{1} {\cal Z}  \left\langle \Psi_i \right| e^{-i \hat H (t-\tau)} \hat a^\dagger_{\bf p} \hat a_{\bf p} e^{-i \hat H \tau} \left| \Psi_i \right\rangle.
\ee
This equation should be solved together with the gap equation \rf{eq:gaps} and \rf{eq:gapcs}. It will fix the value of the chemical potential. 

\subsection{C: Anderson pseudospins}
It is customary when working with Hamiltonians such as those given by \rf{eq:ham3s} or \rf{eq:ham4s} to define the effective spin operators, often called Anderson pseudospins,
\be \hat s^-_{\bf p} = \hat a_{- \bf p} \hat a_{\bf p}, \quad \hat s^+_{\bf p} = \hat a^\dagger_{\bf p}  \hat a^\dagger_{-\bf p}, \quad \hat s^z_{\bf p} = \oh \left( \hat a^\dagger_{{\bf p}} \hat a_{{\bf p}}+ \hat a^\dagger_{-{\bf p}} \hat a_{-{\bf p}} - 1 \right).
\ee
They satisfy the usual su$(2)$ algebra of the angular momentum operators. Note the identification $\hat s^+_{\bf p} = - \hat s^+_{- \bf p}$, $\hat s^-_{\bf p} = - \hat s^-_{- \bf p}$, $\hat s^z_{\bf p} = \hat s^z_{-\bf p}$ which is not essential for what follows. n terms of these, the Hamiltonian \rf{eq:ham4s} can be rewritten as
\be \label{eq:ham6s} \hat H = \sum_{\bf p} \xi_p  \hat s^z - \frac g V \sum_{{\bf p},{\bf q}} p \, q \, \hat s^+_{\bf p} \hat s^-_{\bf q}.
\ee
We can use this Hamiltonian to derive equations of motion of the time-dependent Heisenberg spin operators, with the result
\be  \label{eq:spineom1s} \dot {\hat s}^-_{\bf p} = -  i \xi_{p} {\hat s}^-_{\bf p} -  2 i {\hat \Delta} \, p \, \hat s^z_{\bf p},\quad  \dot {\hat s}^+_{\bf p} =   i \xi_{p} {\hat s}^+_{\bf p}+ 2 i \hat {\bar \Delta} \, p \, {\hat s}^z_{\bf p}, \quad
\dot s^z_{\bf p}  =  i p \left(\hat \Delta \hat s^+_{\bf p} -\hat { \bar \Delta} \hat s^-_{\bf p} \right),
\ee
where
\be \hat \Delta = \frac g V \sum_{\bf p} p \, \hat s^-_{\bf p}, \quad \hat {\bar \Delta} = \frac g V \sum_{\bf p} p \, \hat s^+_{\bf p}.
\ee
Now the key is that operators $\hat \Delta$ and $\hat {\bar \Delta}$ are expressed as a sum over a large number of spins. Therefore, their quantum fluctuations can be neglected and they can be replaced by their expectation values with respect
to the state of the system at the initial moment of time $\left| \Psi_i \right\rangle$, or
\be \Delta = \frac g V \sum_{\bf p} p \, \left\langle \Psi_i \right| \hat s^-_{\bf p} \left| \Psi_i \right\rangle, \quad \bar \Delta = \frac g V \sum_{\bf p} p \,\left\langle \Psi_i \right| \hat s^+_{\bf p} \left| \Psi_i \right\rangle.
\ee
Substituting this into the equations \rf{eq:spineom1s} makes the spin equations of motion linear in operators. As a result, we can replace the remaining operators in these equations of motion by their expectation values, also over the initial state $\left| \Psi_i \right>$,
and obtain fully classical equations of motion
\be  \label{eq:spineom2s} \dot {s}^-_{\bf p} = -  i \xi_{p} { s}^-_{\bf p} -  2 i { \Delta} \, p \, s^z_{\bf p},\quad  \dot { s}^+_{\bf p} =   i \xi_{p} { s}^+_{\bf p}+ 2 i {\bar \Delta} \, p \, {s}^z_{\bf p}, \quad
\dot s^z_{\bf p}  =  i p \left( \Delta s^+_{\bf p} - \bar \Delta s^-_{\bf p} \right),
\ee
where
\be \label{eq:gapcls} \Delta = \frac g V \sum_{\bf p} p \, s^-_{\bf p},  \quad \bar \Delta = \frac g V \sum_{\bf p} p \, s^+_{\bf p}.
\ee
Equations \rf{eq:spineom2s} together with \rf{eq:gapcls} represent nonlinear but exactly solvable (integrable) equations of motion in our problem. Those can in principle be solved to find the classical spins $s^-_{\bf p}$, $s^+_{\bf p}$, $s^z_{\bf p}$ evolving in time.

It is also instructive to rewrite these spins in terms of the amplitudes appearing in the wave function. 
We are interested in time evolution of the states which take the form
\be \label{eq:states} \left| \Psi(\tau) \right\rangle = \prod_{\bf p} \left(u_{\bf p}(\tau) + v_{\bf p}(\tau) \hat s^+_{\bf p} \right) \left| 0 \right\rangle, 
\ee
where $\left|0 \right\rangle$ is the fermion vacuum or the state where all the spins point downwards so that it is annihilated by all the operators $\hat s^-_{\bf p}$. For the initial state $\left| \Psi_i \right\rangle$, $u_{\bf p}$ and $v_{\bf p}$ take
the initial values $u_{i \bf p}$, $v_{i \bf p}$. 

We can derive equations of motion for $u_{\bf p}(\tau)$, $v_{\bf p}(\tau)$ by applying the Hamiltonian \rf{eq:ham6s} to the wave function \rf{eq:states}, which gives
\be \label{eq:1s} i \dot u_{\bf p} = - \xi_p u_{\bf p} -  {\bar \Delta} \, p \,  v_{\bf p}, \ i \dot v_{\bf p}  = \xi_p v_{\bf p} - \Delta \, p \, u_{\bf p},
\ee
where we again apply the same quasiclassical approximation to produce $\Delta$ as given by Eq.~\rf{eq:gapcls}. It is straightforward to relate the classical spins to the amplitudes $u_{\bf p}$, $v_{\bf p}$, by calculating
\be \label{eq:spinam1s} s^+_{\bf p}(\tau) = \left< \Psi(\tau) \right| \hat s^+_{\bf p} \left| \Psi(\tau) \right\rangle = v^*_{\bf p}(\tau) u_{\bf p}(\tau), \ s^-_{\bf p}(\tau) = \left< \Psi(\tau) \right| \hat s^-_{\bf p} \left| \Psi(\tau) \right\rangle = u^*_{\bf p}(\tau) v_{\bf p}(\tau),
\ee
\be \label{eq:spinam2s} s^z_{\bf p}(\tau) = \left< \Psi(\tau) \right| \hat s^z_{\bf p} \left| \Psi(\tau) \right\rangle = \frac{v^*_{\bf p}(\tau) v_{\bf p}(\tau) - u^*_{\bf p}(\tau) u_{\bf p}(\tau)}2.
\ee
Note that following useful relation
\be \label{eq:norms} \sqrt{ s^+_{\bf p} s^-_{\bf p} + \left(s^z_{\bf p} \right)^2 } = \frac 1 2.
\ee
This allows us to rewrite the equation \rf{eq:gapcls} as
\be \label{eq:gapcl1s} \Delta(\tau) = \frac g V \sum_{\bf p} p \, u^*_{\bf p}(\tau) v_{\bf p}(\tau), \quad  \bar \Delta(\tau) = \frac g V \sum_{\bf p} p \, v^*_{\bf p}(\tau) u_{\bf p}(\tau).
\ee
Of course, under the definitions \rf{eq:spinam1s}, \rf{eq:spinam2s}, equations of motion \rf{eq:spineom2s} together with \rf{eq:gapcls} map into \rf{eq:1s} together with \rf{eq:gapcl1s}. Therefore, these two sets of equations represent two equivalent ways of representing the equations of motion in this problem.

Now in our problem as should be clear from the discussion in the previous section, we need to calculate $\Delta$ and $\bar \Delta$ using a different approach,
\be \label{eq:newgaps} \Delta(\tau) = \frac g V \sum_{\bf p} \frac{ p\, \left< \Psi_i \right| e^{-i \hat H (t-\tau)} \hat s^-_{\bf p} e^{-i \hat H \tau} \left| \Psi_i \right\rangle}{ \left\langle \Psi_i \right| e^{-i \hat H t} \left| \Psi_i \right\rangle} , \ \bar \Delta(\tau) = \frac g V \sum_{\bf p}  \frac{ p\, \left< \Psi_i \right| e^{-i \hat H (t-\tau)} \hat s^+_{\bf p} e^{-i \hat H \tau} \left| \Psi_i \right\rangle}{{ \left\langle \Psi_i \right| e^{-i \hat H t} \left| \Psi_i \right\rangle}}.
\ee
To work with these equations we define 
\be \left| \Psi(\tau) \right> = e^{-i \hat H \tau} \left| \Psi_i \right>, \ \left| \tilde \Psi(\tau) \right> = e^{-i  \hat H (\tau - t) } \left| \Psi_i \right>.
\ee
They result in two sets of amplitudes, $u_{\bf p}$, $v_{\bf p}$ as well as $\tilde u_{\bf p}$ and $\tilde v_{\bf p}$. The latter are defined according to Eq.~\rf{eq:states} but with $\Psi$ replaced by $\tilde \Psi$, $u_{\bf p}$ by $\tilde u_{\bf p}$ and
$v_{\bf p}$  by $\tilde v_{\bf p}$.

$\tilde u_{\bf p}$ and $\tilde v_{\bf p}$ satisfy the same equations of motion as their $u_{\bf p}$, $v_{\bf p}$ counterparts, Eq.~\rf{eq:1s} (with the replacement $\bar \Delta \rightarrow \Delta^*$, $\Delta \rightarrow \bar \Delta^*$ \--- recall that $\Delta$ and $\bar \Delta$ are not complex conjugate of each other in the new formalism). However, whereas we have
\be \label{eq:init1s} u_{\bf p}(0) = u_{i \bf p}, \ v_{\bf p}(0) = v_{i \bf p}, 
\ee we have
\be \label{eq:init2s} \tilde u_{\bf p}(t) =  u_{i \bf p}, \ \tilde v_{\bf p}(t) = v_{i \bf p}, 
\ee 
At the same time, we can evaluate the expectation values in Eq.~\rf{eq:newgaps} to produce
\be \label{eq:mixeddeltas} \Delta(\tau)  = \frac g V \sum_{\bf p}  \frac{p\, \tilde u^*_{\bf p}(\tau) v_{\bf p}(\tau)}{2 S_{\bf p}}, \quad \bar \Delta(\tau) = \frac g V \sum_{\bf p}  \frac{ p\, \tilde v^*_{\bf p}(\tau) u_{\bf p}(\tau)}{2 S_{\bf p}}.
\ee
Here we defined
\be \label{eq:sps} 2S_{\bf p} =  \tilde u^*_{\bf p} (\tau) u_{\bf p}(\tau)  + \tilde v^*_{\bf p}(\tau) v_{\bf p}(\tau).
\ee
While the amplitudes on the right hand side of this equation are all dependent on $\tau$, one can check with the equations of motion \rf{eq:1} that $S_{\bf p}$ are independent of $\tau$, while of course they depend on $t$. 
This replaces the earlier equation \rf{eq:gapcl1s} which is not adequate for the purpose of calculating the Loschmidt echo. 

Therefore, we arrive at what appears to be a rather difficult problem. We need to solve the equations of motion \rf{eq:1s} for  $u_{\bf p}$, $v_{\bf p}$ with the initial conditions \rf{eq:init1s} as well as the same equations of motion for their counterparts 
$\tilde u_{\bf p}$, $\tilde v_{\bf p}$ but with the initial conditions \rf{eq:init2s}, all the while keeping $\Delta$ to be given by the Eq.~\rf{eq:mixeddeltas}. 

We can simplify the problem somewhat if we define a new set of variables by generalizing Eqs.~\rf{eq:spinam1s}, \rf{eq:spinam2s} as
\be \label{eq:spinamn1s} s^+_{\bf p}(\tau) = \frac{ \tilde v^*_{\bf p}(\tau) u_{\bf p}(\tau)}{2 S_{\bf p}}, \ s^-_{\bf p}(\tau) =\frac{\tilde u^*_{\bf p}(\tau) v_{\bf p}(\tau)}{2S_{\bf p}},
\ s^z_{\bf p}(\tau) =  \frac{\tilde v^*_{\bf p}(\tau) v_{\bf p}(\tau) - \tilde u^*_{\bf p}(\tau) u_{\bf p}(\tau)}{4 S_{\bf p}}.
\ee
The normalization condition \rf{eq:norms}  still holds for the new variables. In other words, these are new spins -- vectors of fixed length. 
Remarkably, the new spin satisfy the same equations of motion~\rf{eq:spineom2s} as the old ones  and, moreover, the gap equations also retain their form \rf{eq:gapcls}.  So this approach has the advantage of keeping the familiar equations of motion. 
The difficulty is now that the initial conditions \rf{eq:init1s} and \rf{eq:init2s} must now be reworked in terms of these spins. One can check that this gives
 \be \label{eq:newboundarys} \frac{1+2 s^z_{\bf p}(0)}{  2 s^+_{\bf p}(0)} =  \frac{v_{i \bf p}}{u_{i \bf p}}, \  \frac{1+2 s^z_{\bf p}(t)}{  2 s^-_{\bf p}(t)} =  \frac{v^*_{i \bf p}}{u^*_{i \bf p}}.
\ee
Therefore, the goal is now to solve the equations of motion \rf{eq:spineom2s} with \rf{eq:gapcls}, while imposing the boundary conditions \rf{eq:newboundarys}. One should also note that for these new spins $s^+_{\bf p}$ is not
equal to the complex conjugate of $s^-_{\bf p}$, and $s^z_{\bf p}$ is not necessarily real. 

Looking for complex solutions of the equations of motion with the boundary conditions \rf{eq:newboundarys} generally remains an open problem.

\subsection{D: Classical echo}
In calculating Loschmidt echo we need to compute
\be \label{eq:echo2s} {\cal Z} = \left< \Psi_i \right| e^{-i \hat H t} \left| \Psi_i \right\rangle = \prod_{\bf p} 2 S_{\bf p}.
\ee
Here $S_{\bf p}$ are defined in Eq.~\rf{eq:sps}. 

However, if we were not careful, we could have instead computed 
\be \left| \Psi(t) \right\rangle  = e^{- i \hat H t} \left| \Psi_i \right\rangle
\ee using the conventional approach of solving equations \rf{eq:1s} together with \rf{eq:gapcl1s}. We could then calculate
\be \left< \Psi_i \Big| \Psi(t) \right\rangle = \prod_{\bf p} \left( u^*_{i \bf p} u_{\bf p}(t) + v^*_{i \bf p} v_{\bf p}(t) \right),
\ee
The right hand side of this equation appears to be superficially similar to the right hand side of the equation \rf{eq:echo2s} if $\tau=t$ is used in the definition of $S_{\bf p}$, Eq.~\rf{eq:sps}. However, importantly in evaluating the amplitudes $u_{\bf p}$,
$v_{\bf p}$, Eq.~\rf{eq:gapcl1s} is used instead of Eq.~\rf{eq:mixeddeltas}. This in fact was done in the earlier publication of some of the authors, Ref.~\onlinecite{Rylands2021}. 

One could ask about the meaning of the resulting quantity. We would like to present arguments that it is equivalent to the {\sl classical echo} introduced in Ref.~\onlinecite{Prosen2004}. 

Consider the quantity
\be {\cal L}(t) = \left| \left\langle \Psi_i \Big| \Psi(t) \right\rangle \right|^2.
\ee
It is a matter of simple algebra to show~\cite{Rylands2021} that
\be
\mathcal{L}(t)=\prod_{\bf p} \mathcal{L}_{\bf p}(t),
\label{echo}
\ee
where
\be
 \mathcal{L}_{\bf p}(t)=\frac{1}{2}+ 2 {\bf s}_{\bf p}(0)\cdot {\bf s}_{\bf p} (t),
 \label{lpp}
\ee
with the spins defined according to Eq.~\rf{eq:spinam1s}, \rf{eq:spinam2s}. 

Classical Loschmidt echo is defined as~\cite{Prosen2004}
\be
\mathcal{L}_\mathrm{cl}(t)=\int d{\bf x}\, \rho({\bf x}, 0) \rho({\bf x}, t),
\label{ekho_gens}
\ee
where the integral is over the phase space ${\bf x}$ and $\rho({\bf x}, t)$ is the phase space density of a classical system. The classical counterpart of spin is angular momentum (classical spin) and the corresponding phase space density is the Wigner function of spin-$\frac{1}{2}$~\cite{Davis2021},
\be
\rho({\bf k}, t=0)=\frac{1}{2}+ \sqrt{3} \, {\bf k}\cdot{\bf s},
\label{wigners}
\ee
where ${\bf k}$ is a unit vector representing the phase space of the spin ${\bf s}$. We choose  to parametrize ${{\bf k}}$ by its spherical angles $(\theta, \phi)$, i.e.,
\be
{\bf k} =[ \sin\theta\cos\phi, \sin\theta\sin\phi, \cos\theta].
\ee 
Eq.~\rf{wigners} is a distribution of initial conditions  for a classical spin that rotates  according to  classical equations of motion \rf{eq:spineom2s}. Each initial condition [initial direction ${\bf k}$ of the  classical spin] moves to a new point ${\bf k}'$ in the phase space in time $t$ thus generating
the evolution of the Wigner function in time.  

It is sufficient to do the calculation for one spin ${\bf s}_{\bf p}$ due to the product nature of \rf{echo}. ${\bf s}_{\bf p}(0)$ and ${\bf s}_{\bf p}(t)$ are related by a rotation. Since we did not pick coordinate axis until now, we can choose them arbitrarily without  loss of generality. Let the axis of rotation be the $z$-axis and  ${\bf s}_{\bf p}(0)$ be in the $xz$-plane with spherical coordinates
$\theta=\theta_0$ and $\phi=0$.  In this coordinate system we also have
\be
{\bf s}_{\bf p}(0)\cdot {\bf s}_{\bf p}(t) =\frac{1}{4} \cos^2\theta_0+\frac{1}{4}\sin^2\theta_0\cos\alpha,
\ee
and
\be
{\bf s}_{\bf p}(0)=[  \sin\theta_0, 0, \cos\theta_0]/2.
\ee

The phase space point $(\theta, \phi)$ evolves to $(\theta, \phi+\alpha)$ under this rotation, where $\alpha$ is the angle of rotation and
the vector ${\bf k}$ evolves into 
\be
{\bf k}' =[ \sin\theta\cos(\phi+\alpha), \sin\theta\sin(\phi+\alpha), \cos\theta],
\ee 
and
\be
\rho(\theta, \phi, t)=\frac{1}{2}+\sqrt{3} \, {\bf k}'\cdot{\bf s}(0).
\label{wigner's}
\ee
The integration in Eq.~\rf{ekho_gens} should be understood as integration (averaging) over the sphere, i.e., 
\be
\int d{\bf x} = \frac{1}{4\pi} \int_{-1}^1 d\cos\theta \int_0^{2\pi} d\phi
\ee
Note that Wigner function~\rf{wigners} is  normalized to one with this integration measure. Using this definition of $\int d{\bf x}$
in Eq.~\rf{ekho_gens} with $\rho({\bf x}, 0)$ and  $\rho({\bf x}, t)$ from Eqs.~\rf{wigners} and \rf{wigner's} and  the above expressions for
${\bf s}$, ${\bf k}$, and ${\bf k}'$, we obtain
\be
\mathcal{L}_\mathrm{cl}= \frac{1}{4}+{\bf s}_{\bf p}(0)\cdot {\bf s}_{\bf p}(t). 
\ee
This coincides with $\mathcal{L}_{\bf p}$ in Eq.~\rf{lpp} up to a nonessential factor of 2.

\end{document}